\documentclass[amsmath,amssymb,superscriptaddress,aps,prl,twocolumn,reprint]{revtex4-1}

\usepackage{graphicx}
\usepackage{amsfonts,amsmath,amssymb}

\begin{document}

\title{A sparse-sampling approach for the fast computation of matrices:
application to molecular vibrations}


\author{Jacob N. Sanders}
\author{Xavier Andrade}
\author{Al\'{a}n Aspuru-Guzik}
\email{aspuru@chemistry.harvard.edu}
\affiliation{Department of Chemistry and Chemical Biology, Harvard
  University, 12 Oxford Street, Cambridge, MA 02138, United States}



\begin{abstract}
This article presents a new method to compute matrices from numerical
simulations based on the ideas of sparse sampling and compressed
sensing. The method is useful for problems where the determination of
the entries of a matrix constitutes the computational bottleneck. We
apply this new method to an important problem in computational
chemistry: the determination of molecular vibrations from electronic
structure calculations, where our results show that the overall scaling
of the procedure can be improved in some cases. Moreover, our method
provides a general framework for bootstrapping cheap low-accuracy
calculations in order to reduce the required number of expensive
high-accuracy calculations, resulting in a significant 3$\times$
speed-up in actual calculations.
\end{abstract}

\maketitle

%
%

\section{Introduction}

Matrices are one of the most fundamental objects in the mathematical description of nature, and as such they are ubiquitous in every area of science. For example, they arise naturally in linear response theory as the first term in a multidimensional Taylor series, encoding the response of each component of the system to each component of the stimulus.  Hence, in many scientific applications, matrices contain the essential information about the system being studied.

Despite their ubiquity, the calculation of matrices often requires considerable computational effort.  Returning to the linear response theory example, it might be necessary to individually calculate the response of every component of the system to every component of the stimulus and, depending on the area of application, each individual computation may itself be quite expensive.  The overall expense stems from the fact that evaluating a matrix of dimension \(N\times M\) requires, in principle, the individual evaluation of \(N\times M\) elements.  But this does not always have to be the case.

For example, if we know \emph{a priori} the eigenvectors of a \(N\times N\) diagonalizable matrix, then we can obtain the full matrix by only calculating the \(N\) diagonal elements.  Similarly, a sparse matrix, which contains many zero elements, can be evaluated by calculating only the non-zero elements, if we know in advance where such elements are located.  In this article, we present a general approach that can produce a considerable reduction in the cost of constructing a matrix in many scientific applications by substantially reducing the number of elements that need to be calculated.  

The key numerical procedure of our approach is a method to cheaply recover sparse matrices with a cost that is essentially proportional to the number of non-zero elements. The matrix reconstruction procedure is based on the increasingly popular compressed sensing approach~\cite{Candes2006,Donoho2006,Candes2008}, a state-of-the-art signal processing technique developed to minimize the amount of data that needs to be measured to reconstruct a sparse signal.

The use of compressed sensing and sparse sampling methods for scientific development has been dominated by experimental applications~\cite{Hu_2009,Doneva_2010,Gross_2010,Kazimierczuk_2011,Holland_2011,Shabani_2011,Zhu_2012,Sanders2012,Song_2012,August_2013,Xu_2013}. However compressed sensing is also becoming a tool for theoretical applications~\cite{Andrade2012b,Almeida_2012,Schaeffer_2013,Nelson_2013,Markovich2014,Flammia_2012,Baldwin_2014}. In particular, in previous work we have shown that compressed sensing can also be used to reduce the amount of computation in numerical simulations~\cite{Andrade2012b}.

In this article, we apply compressed sensing to the problem of computing matrices.  This method has two key properties.  First, the cost of the procedure is quasi-linear with the size of the number of non-zero elements in the matrix, without the need to know \emph{a priori} the location of the non-zero elements.  Second, the rescontruction is exact.  Furthermore, the utility of the method extends beyond the computation of \emph{a priori} sparse matrices.  In particular, the method suggests a new computing paradigm in which one develops methods to find a basis in which the matrix is known or suspected to be sparse, based on the characteristics and prior knowledge of the matrix, and then afterwards attempts to recover the matrix at lower computational cost.

To demonstrate the power of our approach, we apply these ideas to an important problem in quantum chemistry: the determination of the vibrational modes of molecules from electronic structure methods.  These methods require the calculation of the matrix of the second derivatives of the energy with respect to the nuclear displacements, known as the force-constant or Hessian matrix.  This matrix is routinely obtained in numerical simulations by chemists and physicists, but it is relatively expensive to compute when accurate quantum mechanical methods are used.  Our application shows that we can exploit the sparsity of the matrix to make important improvements in the efficiency of this calculation.  At the same time, our method provides a general framework for bootstrapping cheap low-accuracy calculations to reduce the required number of expensive high-accuracy calculations, something which previously was not possible to do in general.

We begin by discussing how compressed sensing makes it practical to take a new approach for the calculation of matrices based on finding strategies to make the matrix sparse.  
Next, we introduce the mathematical foundations of the method of compressed sensing and apply them to the problem of sparse matrix reconstruction. This is the numerical tool that forms the foundation of our approach. Finally, we illustrate these new ideas by applying them to the problem of obtaining molecular vibrations from quantum mechanical simulations.

\section{Finding a sparse description of the problem}

The first step in our approach is to find a representation for the problem where the matrix to be calculated is expected to be sparse.  In general, finding this \textit{sparsifying basis} is specific to each problem and ranges from trivial to quite complex; it has to do with the knowledge we have about the problem or what we expect about its solution.

Leveraging additional information about a problem is an essential concept in compressed sensing, but it is also a concept that is routinely exploited in numerical simulations. For example, in quantum chemistry it is customary to represent the orbitals of a molecule in a basis formed by the orbitals of the atoms in the molecule~\cite{Szabo1996}, which allows for an efficient and compact representation and a controlled discretization error.  This choice comes from the notion that the electronic structure of the molecule is roughly described by ``patching together'' the electronic structure of the constituent atoms.

An ideal basis in which to reconstruct a matrix is the basis of its eigenvectors, or \textit{eigenbasis}, as this basis only requires the evaluation of the diagonal elements to obtain the whole matrix.  Of course, finding the eigenbasis requires knowing the matrix in the first place, so reconstructing a matrix in its eigenbasis is not useful for practical purposes.  However, in many cases it is possible to obtain a set of reasonable approximations to the eigenvectors (an idea which also forms the basis of perturbation theory in quantum mechanics). The approximate eigenbasis probably constitutes a good sparsifying basis for many problems, as we expect the matrix to be diagonally dominant, with a large fraction of the off-diagonal elements equal to zero or at least quite small.

Since the determination of an approximate eigenbasis depends on the specific problem at hand, a general prescription is difficult to give.  Nevertheless, a few general ideas could work in many situations.  For example, in iterative or propagative simulations, results from previous iterations or steps could be used to generate a guess for the next step.  Alternatively, cheap low-accuracy methods can be used to generate a guess for an approximate eigenbasis.  In this case, the procedure we propose provides a framework for bootstraping the results of a low-cost calculation in order to reduce the required number of costly high-accuracy calculations. This last strategy is the one we apply to the case of molecular vibrations.

What makes looking for sparsifying basis attractive, even at some computational cost and code-complexity overhead, are the properties of the recovery method.  First, the cost of recovering the matrix is roughly proportional to its sparsity.  Second, the reconstruction of the matrix is always exact up to a desired precision; even if the sparsifying basis is not a good one, we eventually converge to the correct result.  The penalty for a bad sparsifying basis is additional computation, which in the worst case makes the calculation as costly as if compressed sensing were not used at all.  This feature implies that the method will almost certainly offer some performance gain.

There is one important qualification to this performance gain. For some matrices, there is a preferred basis in which the matrix is cheaper to compute, and the extra cost of computing its elements in a different basis might offset the reduction in cost offered by compressed sensing.

\section{Compressed sensing for sparse matrices}

Once a sparse representation for the matrix is known, the numerical core of our method for the fast computation of matrices is the application of compressed sensing to calculate sparse matrices without knowing \textit{a priori} where the non-zero elements are located. Related work has been presented in the field of compressive principal component pursuit~\cite{Baraniuk2011,Candes2011,Zhou2010,Wright2013}, which focuses on reconstructing matrices that are the sum of a low-rank component and a sparse component.  Our work instead outlines a general procedure for reconstructing any sparse matrix by measuring it in a different basis.

Suppose we wish to recover a \(N \times N\) matrix \(A\) known to be sparse in a particular orthonormal basis \(\{\psi_i\}\) (for simplicity we restrict ourselves to square matrices and orthonormal bases).  Without any prior knowledge of where the \(S\) non-zero elements of \(A\) are located, it might appear that we need to calculate all \(N^2\) elements, but this is not the case.

In a different orthonormal basis \(\{\phi_i\}\), the matrix \(A\) has a second representation \(B\) given by 
\begin{equation}
\label{eq:cob_matrix}
B = PAP^{T}\ ,
\end{equation}
where \(P\) is the orthogonal change-of-basis matrix from the basis \(\{\psi_i\}\) to the basis \(\{\phi_i\}\).  Note that in general \(B\) is not sparse.

\begin{figure}
\begin{center}
\includegraphics[width=1\columnwidth]{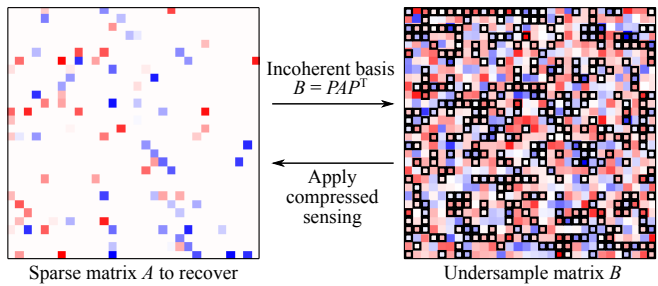}
\caption{\label{fig:NumericsScheme} General scheme for the recovery of a sparse matrix \(A\) via compressed sensing.  Rather than sampling \(A\) directly, the key is to sample the matrix \(B\) which corresponds to \(A\) expressed in an different (known) basis.  Recovery of \(A\) from the undersampled entries of \(B\) proceeds via compressed sensing by solving eq.~\ref{eq:bpdn}.}
\end{center}
\end{figure}

If we regard \(A\) and \(B\) as \(N^2\)-element \emph{vectors}, it is easy to see that the change-of-basis transformation from \(A\) to \(B\) given by eq.~\ref{eq:cob_matrix} is linear.  This fact enables us to use the machinery of compressed sensing to reconstruct the full matrix \(A\) by sampling only \textit{some} of the entries of \(B\). The reconstruction is done by solving the basis pursuit (BP) problem \cite{Berg2008},
\begin{equation}
  \label{eq:bpdn}
  \min_{A} ||A||_1 \quad \textrm{subject to} \quad (PAP^T)_{ij} = B_{ij}\quad \forall
\ i,j \in W \ ,
\end{equation}
where the 1-norm is considered as a \emph{vector} norm (\(||A||_1 = \sum_{i,j} \left|A_{ij}\right|\)), and \(W\) is a set of randomly chosen entries in matrix \(B\).  This approach to matrix reconstruction is illustrated in Fig.~\ref{fig:NumericsScheme}.

The size of the set \(W\), a number that we call \(M\), is the number of matrix elements of \(B\) that are sampled. \(M\) determines the quality of the reconstruction of \(A\).  From compressed sensing theory we can find a lower bound for \(M\) as a function of the sparsity of \(A\) and the change-of-basis transformation.

One important requirement for compressed sensing is that the sparse basis \(\{\psi_i\}\) for \(A\) and the measurement basis \(\{\phi_i\}\) for \(B\) should be \textit{incoherent}, meaning that the maximum overlap between any vector in \(\{\psi_i\}\) and any vector in \(\{\phi_i\}\)
\begin{equation}
\label{eq:coherence}
\mu = \sqrt{N} \max_{i,j} \langle \psi_i | \phi_j \rangle
\end{equation}
should be as \emph{small} as possible (in general \(\mu\) ranges from 1 to \(\sqrt{N}\)).  Intuitively, this incoherence condition means that the change-of-basis matrix \(P\) should thoroughly scramble the entries of \(A\) to generate \(B\). 

It can be proven~\cite{Candes2008} that the number of entries of \(B\) which must be measured in order to fully recover \(A\) by solving the BP problem in eq.~\ref{eq:bpdn} scales as
\begin{equation}
\label{eq:csscaling}
M \propto \mu^2 S \log N^2\ .
\end{equation}
This scaling equation encapsulates the important aspect of compressed sensing: if a proper measurement basis is chosen, the number of entries which must be measured scales linearly with the sparsity of the matrix and only depends weakly on the full size of the matrix. For the remainder of this paper, we always choose our measurement basis vectors to be the discrete cosine transform (DCT) of the sparse basis vectors, for which the parameter \(\mu\) is equal to \(\sqrt{2}\).

In order to study the numerical properties of the reconstruction method we performed a series of numerical experiments. We generate \(100 \times 100\) matrices of varying sparsity with random values drawn uniformly from the interval \([-1,1]\) and places in random locations in the matrix. Matrix elements were then sampled in the DCT measurement basis, and an attempt was made to recover the original sparse matrix by solving the compressed sensing basis pursuit problem in eq.~\ref{eq:bpdn}.

\begin{figure}
\begin{center}
\includegraphics[width=1\columnwidth]{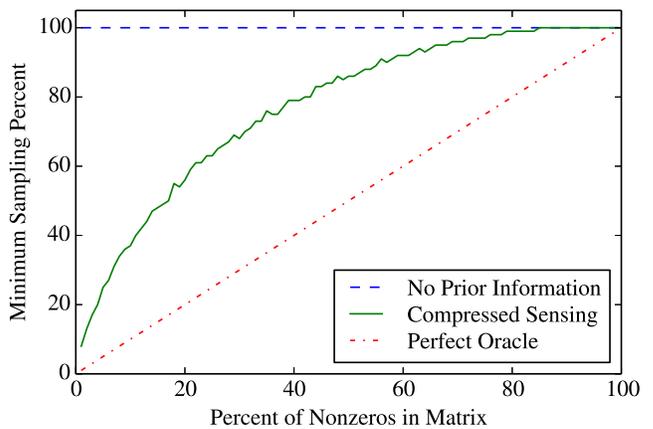}
\caption{\label{fig:NumericsResults} Percent of entries that must be sampled for accurate recovery of a matrix as a function of sparsity. Comparison between compressed sensing and two limiting cases: ``no prior knowledge'' of sparsity and the ``perfect oracle'' who reveals where all non-zero entries are located.  Each point on the compressed sensing curve is an average of ten different randomizations.  The accuracy criterion is a relative error in the Frobenius norm smaller than \(10^{-7}\). }
\end{center}
\end{figure}

Fig.~\ref{fig:NumericsResults} illustrates the percent of matrix elements had to be sampled for accurate recovery of the sparse matrix compared with other recovery approaches. If no prior knowledge of a matrix is used for its recovery, then one simply measures each entry; this is the current paradigm in many scientific applications.  If one knows exactly where the non-zeros in a sparse matrix are located, one can simply measure those elements. Compressed sensing interpolates between these two extremes: it provides a method for recovering a sparse matrix when the locations of the non-zeros are not known in advance. Although this lack of knowledge comes with a cost, the recovery  is still consideably cheaper than measuring the entire matrix.

\section{Application: molecular vibrations}

Calculating the vibrations of a molecule, both the frequencies and associated normal modes, is one of the most ubiquitous tasks in computational chemistry~\cite{Wilson1955}. Integrated into nearly all computational-chemistry packages, molecular vibrations are routinely computed by theoretical and experimental chemists alike.  Chemists routinely optimize molecular geometries to find minimal energy conformations; computing and confirming the positivity of all vibrational frequencies is the standard method of assuring that a local minimum has been found.  Another common task is to find the transition state for a proposed reaction: here it is also necessary to compute the molecular vibrations to find one mode with an imaginary frequency, confirming the existence of a local maximum along the reaction coordinate~\cite{Cramer2004}.  Despite the centrality of molecular vibrations in computational chemistry, it remains one of the most expensive computations routinely performed by chemists.

The core of the technique lies in calculating the matrix of the mass-weighted second derivatives of the energy with respect to the atomic positions
\begin{equation}
\label{eq:hessian}
H_{Ai,Bj} = \frac{1}{\sqrt{M_A M_B}}\frac{\partial^2E(\vec{R}^1,\ldots,\vec{R}^N)}{\partial R^A_i \partial R^B_j}\,
\end{equation}
where \(E(\vec{R}^1,\ldots,\vec{R}^N)\) is the ground-state energy of the molecule, \(R^A_i\) is coordinate \(i\) of atom \(A\), and \(M_A\) is its mass.  Hence, the Hessian is a real \(3N \times 3N\) matrix where \(N\) is the number of atoms in the molecule.  When the molecule is in a minimum energy conformation, the eigenvectors of the Hessian correspond to the vibrational modes of the molecule, and the square root of the eigenvalues correspond to the vibrational frequencies~\cite{Cramer2004}.

Our goal, therefore, is to understand how our approach can reduce the cost of computing the Hessian matrix of a molecule.  We achieve this understanding in two complementary ways.  First, for a moderately-sized molecule, we outline and perform the entire numerical procedure to show in practice what kinds of speed-ups may be obtained.  Second, for large systems, we investigate the ability of compressed sensing to improve how the cost of computing the Hessian scales with the number of atoms.

Calculating the Hessian requires a method for obtaining the energy of a given nuclear configuration. There exist many methods to chose from, which offer a trade-off between accuracy and computational cost.  Molecular mechanics approaches, which model the interactions between atoms via empirical potentials~\cite{Cramer2004}, are computationally cheap for systems of hundreds or thousands of atoms, while more accurate and expensive methods explicitly model the electronic degrees of freedom at some level of approximated quantum mechanics, such as methods based on density functional theory (DFT)~\cite{Hohenberg_1964,Kohn_1965} or wavefunction methods~\cite{Szabo1996}.  We focus on these quantum mechanical approaches, since in that type of calculations the computation time is dominated by the calculation of the elements of the Hessian matrix, making it an ideal application for our  matrix-recovery method.

\begin{figure}
\begin{center}
\includegraphics[width=1\columnwidth]{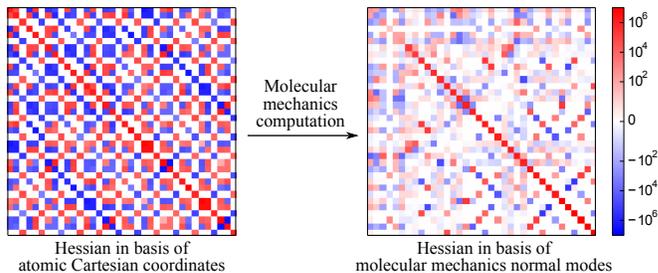}
\caption{\label{fig:HessianScheme}  The quantum mechanical Hessian of benzene in the basis of atomic Cartesian coordinates (on the left) and in the basis of molecular mechanics normal modes (on the right).  Since the molecular mechanics normal modes form approximate eigenvectors to the true quantum mechanical Hessian, the matrix on the right is sparse (close to diagonal) and therefore well-suited to recovery via compressed sensing.}
\end{center}
\end{figure}

To recover a quantum mechanical Hessian efficiently with compressed sensing, we need to find a basis in which the matrix is sparse. While we might expect to the Hessian to have some degree of sparsity in the space of atomic Cartesian coordinates, especially for large molecules, we have found that it is possible to find a better basis. The approach we take is to use a basis of approximated eigenvectors generated by a molecular mechanics computation that provide a cheap approximation to the eigenvectors of the quantum mechanical Hessian. This is illustrated in Fig.~\ref{fig:HessianScheme} for the case of the benzene molecule (\(\textrm{C}_{6}\textrm{H}_{6}\)). The figure compares the quantum mechanical Hessian in the basis of atomic Cartesian coordinates with the same matrix in the approximate eigenbasis obtained via an auxiliary molecular mechanics computation.  As the figure shows, the matrix in the molecular mechanics basis is much sparser, and is therefore better suited to recovery via compressed sensing.

The second derivatives of the energy required for the Hessian, eq.~\ref{eq:hessian}, can be calculated either via finite differences, generating what are known as \textit{numerical} derivatives, or using perturbation theory, generating so-called \textit{analytical} derivatives~\cite{Pulay1969,Jorgensen1983,Baroni2001,Kadantsev2005}. 

A property of the calculations of the energy derivatives is that the numerical cost does not depend on the direction they are calculated. This can be readily seen in the case of finite differences, as the cost of calculating \(E(\vec{R}^1,\ldots,\vec{R}^j+\Delta^j,\ldots,\vec{R}^N)\) is essentially the same as computing \(E(\vec{R}^1 + \Delta^1,\ldots,\vec{R}^j+\Delta^j,\ldots,\vec{R}^N + \Delta^N)\). As discussed previously, this ability to compute matrix elements at a comparable numerical cost in any desired basis is an essential requirement of our method.


A second property of both numerical and analytical derivatives that appears in variational  quantum chemistry formalisms like DFT or Hartree-Fock is that each calculation yields a full column of the Hessian, rather than a single matrix element. Again, this is easy to see in finite difference computations. We can write the second derivative of the energy as a first derivative of the force
\begin{equation}
\label{eq:force}
\frac{\partial^2E(\vec{R}^1,\ldots,\vec{R}^N)}{\partial R^A_i \partial R^B_j} =
- \frac{\partial F^B_j(\vec{R}^1,\ldots,\vec{R}^N)}{\partial R^A_i} \ .
\end{equation}
By the Hellman-Feynman theorem~\cite{Hellmann1937,Feynman1939}, a single energy calculation yields the forces acting over \emph{all} atoms, so the evaluation of eq.~\ref{eq:force} by finite differences for fixed \(A\) and \(i\) yields the derivatives for all values of \(B\) and \(j\), a whole column of the Hessian.  An equivalent result holds for analytic derivatives obtained via perturbation theory~\cite{Baroni2001,Kadantsev2005}. Thus, our compressed sensing procedure for this particular application focuses on measuring random columns of the quantum mechanical Hessian rather than individual random entries.

The full compressed sensing procedure applied to the calculation of a quantum mechanical Hessian is therefore implemented as the following:

\begin{enumerate}
\item Calculate approximate vibrational modes using molecular mechanics.
\item Transform the approximate modes using the DCT matrix.
\item Randomly select a few of the transformed modes.
\item Calculate the energy second derivatives along these random modes to yield a set of random columns of the quantum mechanical Hessian.
\item Apply compressed sensing to rebuild the full quantum mechanical Hessian in the basis of approximate vibrational modes.
\item Transform the full quantum mechanical Hessian back into the atomic coordinate basis.
\item Diagonalize the quantum mechanical Hessian to obtain the vibrational modes and frequencies.
\end{enumerate}

\begin{figure}
\begin{center}
\includegraphics[width=1\columnwidth]{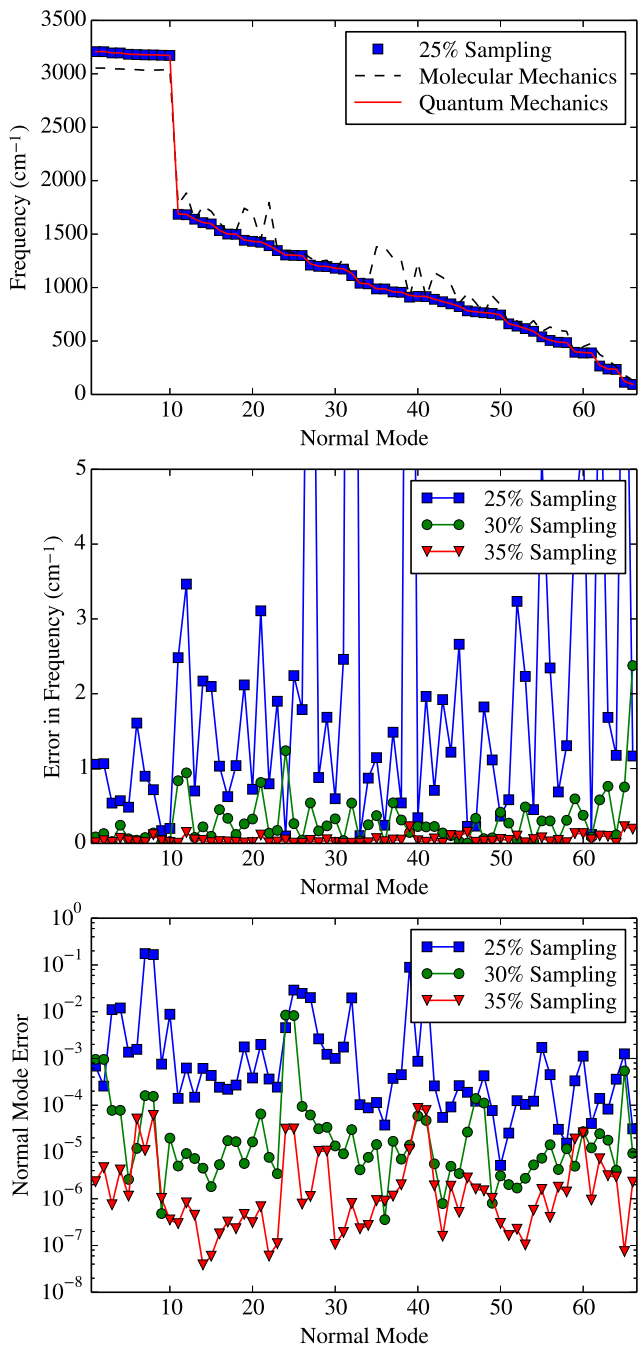}
\caption{\label{fig:Anthracene} Results of applying our compressed sensing procedure to the vibrational modes and frequencies of anthracene.  (Top) Even by sampling only 25\% of the quantum mechanical Hessian, the vibrational frequencies obtained via compressed sensing converge to those of the true quantum mechanical Hessian. (Middle) Error in vibrational frequencies for different extents of undersampling. When only 30\% of the columns are sampled, the maximum error in frequency is within 3 cm\(^{-1}\), and with 35\% sampling, the recovery is essentially exact.  (Bottom) Error in vibrational normal modes for different extents of undersampling on a logarithmic scale; the error is calculated as one minus the overlap (dot product) between the exact quantum mechanical normal mode and the normal mode obtained via compressed sensing. Once 30\% of the columns are sampled, the normal modes are recovered to within 1\% accuracy.}
\end{center}
\end{figure}

Fig.~\ref{fig:Anthracene} illustrates the results of applying our Hessian recovery procedure to anthracene (\(\textrm{C}_{14}\textrm{H}_{10}\)), a moderately-sized polyacene consisting of three linearly fused benzene rings.  The top panel illustrates the vibrational frequencies obtained by the compressed sensing procedure outlined above for different extents of undersampling of the true quantum mechanical Hessian.  Even sampling only 25\% of the columns yields vibrational frequencies that are close to the true quantum mechanical frequencies, and much closer than the molecular mechanics frequencies.  The middle panel illustrates the error in the vibrational frequencies from the true quantum mechanical frequencies.  Sampling only 30\% of the columns gives rise to a maximum frequency error of less than 3 cm\(^{-1}\), and sampling 35\% of the columns yields nearly exact recovery.  The bottom panel illustrates the error in the normal modes.  Once again, sampling only 30\% of the columns gives accurate recovery of all vibrational normal modes to within 1\%.  In short, our compressed sensing procedure applied to anthracene reduces the number of expensive quantum mechanical computations by a factor of three. The additional cost of the molecular mechanics computation and the compressed sensing procedure, which take a few seconds, is negligible in comparison with this reduction in computational time in the computation of the Hessian that for anthracene takes on the order of hours.

Having shown that our compressed sensing procedure results in a 3\(\times\) speed-up for a moderately-sized organic molecule, we now move to larger systems and investigate how the cost of computing the Hessian scales with the number of atoms.  In the absence of compressed sensing, if the entries of the Hessian must be calculated independently, the cost of calculating the Hessian would scale as \(O(N^2) \times OE\), where \(OE\) is the cost of computing the energy of a given nuclear configuration (the cost of analytical and numerical derivatives usually have the same scaling). For example, for a DFT-based calculation, \(OE\) is typically \(O(N^3)\).  However, since many quantum mechanical methods obtain the Hessian one column at a time, only \(O(N)\) calculations are required, so the scaling is improved to \(O(N)\times OE\).

How does compressed sensing alter this scaling?  From eq.~\ref{eq:csscaling}, the number of matrix elements needed to recover the Hessian via compressed sensing scales as \(O(S \log N)\), where \(S\) is number of non-zero elements in the Hessian, so the net scaling is \(O(S \log N) \times OE\).  By obtaining the Hessian one column at a time, we expect the net scaling to improve to \(O(S/N \log N) \times OE\). However, we should note that eq.~\ref{eq:csscaling} is only valid in principle for the random sampling of elements, and it is not necessarily valid for a random column sampling. This scaling result illustrates the critical importance of recovering the Hessian in a sparse basis, with \(S\) as small as possible. So what is the smallest \(S\) that can reasonably be achieved?

For many large systems, the Hessian is already sparse directly in the basis of atomic Cartesian coordinates.  Since the elements of the Hessian are \emph{partial} second derivatives of the energy with respect to the positions of two atoms, only direct interactions between the two atoms, with the positions of all other atoms held fixed, must be taken into account.  For most systems we expect that this direct interaction has a finite range or decays strongly with distance.   Note that this does not preclude collective vibrational modes, which can emerge as a result of ``chaining together'' direct interactions between nearby atoms.

If we assume that a system has a finite range interaction between atoms, and since each atom has an approximately constant number of neighbors, irrespective of the total number of atoms in the molecule, the number of non-zero elements in a single column of the Hessian should be constant.  Hence, for large molecules, the sparsity \(S\) of the Hessian would scale linearly with the number of atoms \(N\).  Putting this result into \(O(S/N \log N) \times OE\) yields a best-case scaling of \(O\left(\log N \right)\times OE\), which is a significant improvement over the original \(O(N)\times OE\) in the absence of compressed sensing.

\begin{figure}
\begin{center}
\includegraphics[width=0.98\columnwidth]{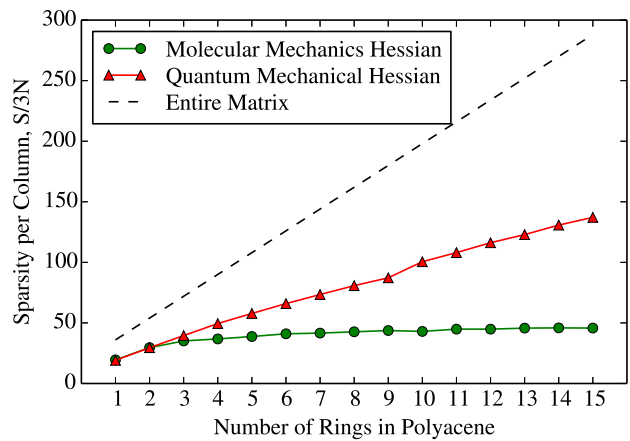}
\caption{\label{fig:HessSparsity} Average sparsity per column (\(S/3N\)) of molecular mechanics and quantum mechanical Hessians in the basis of atomic coordinates for the series of polyacenes.  (An entry in the Hessian is considered nonzero if it is greater than 10 \((\mathrm{cm}^{-1})^2\), which is roughly six orders of magnitude smaller than the largest entry.)  In the molecular mechanics Hessians, the average sparsity per column is roughly constant with the size of the molecule, because each atom has a roughly constant number of neighbors regardless of the size of the entire molecule.}
\end{center}
\end{figure}

To study the validity of our scaling results we have performed numerical calculations on a series of polyacene molecules, which are aromatic compounds made of linearly fused benzene rings.  For polyacenes ranging from 1 to 15 rings, Fig.~\ref{fig:HessSparsity} illustrates the average number of non-zeros per column in the Hessian matrices obtained via molecular mechanics and quantum mechanical calculations in the basis of atomic coordinates.  In the molecular mechanics Hessians, the average sparsity per column approaches a constant value as the size of the polyacene increases, consistent with each atom having direct interaction with a constant number of other atoms.

\begin{figure}
\begin{center}
\includegraphics[width=1\columnwidth]{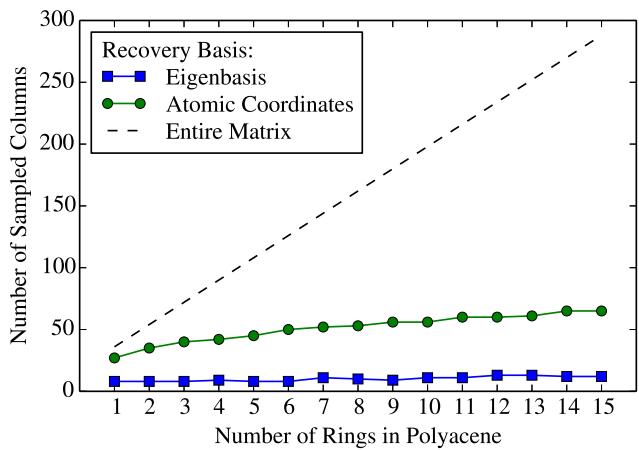}
\caption{\label{fig:MMNColumnsVsNRings} Number of columns which must be sampled as a function of the number of rings in the polyacene to achieve a relative Frobenius norm error less than $10^{-3}$ in the recovered molecular mechanics Hessian.  Legend entries indicate the (sparse) recovery basis, and columns are always sampled in the DCT basis with respect to the recovery basis.  (Relative error is measured by averaging over ten different trials which sample different sets of random columns.)}
\end{center}
\end{figure}

Since the molecular mechanics Hessians illustrate the best-case scenario in which the sparsity \(S\) scales linearly with the number of atoms \(N\), we attempted to recover these Hessians directly in the basis of atomic coordinates via the compressed sensing procedure we have outlined by sampling columns in the DCT basis. Fig.~\ref{fig:MMNColumnsVsNRings} illustrates the number of columns which must be sampled to recover the Hessians to within a relative error of \(10^{-3}\) as a function of the size of the polyacene.  Far fewer than the total number of columns in the entire matrix need to be sampled.  Even more attractive is the fact that the number of columns grows quite slowly with the size of the polyacene, consistent with the best-case \(O\left(\log N \right)\times OE\) scaling result obtained above.  This result indicates that our compressed sensing approach is especially promising for the calculation of Hessian matrices for large systems. For comparison, we also recovered the Hessians in their sparsest possible basis, which is their own eigenbasis. This procedure is not practical for actual calculation since it requires knowing the entire Hessian beforehand, but it shows the best-case scenario and illustrates how the compressed sensing procedure can be improved further if an appropriate sparsifying transformation is known.

While the recovery of molecular mechanics Hessians provides a clear illustration of the scaling of our compressed sensing procedure, molecular mechanics matrix elements are not expensive to compute in comparison with rest of the linear algebra operations required to diagonalize the Hessian.  Hence, from a computational standpoint, the real challenge is to apply our procedure to the computation of quantum mechanical Hessians.

\begin{figure}
\begin{center}
\includegraphics[width=1\columnwidth]{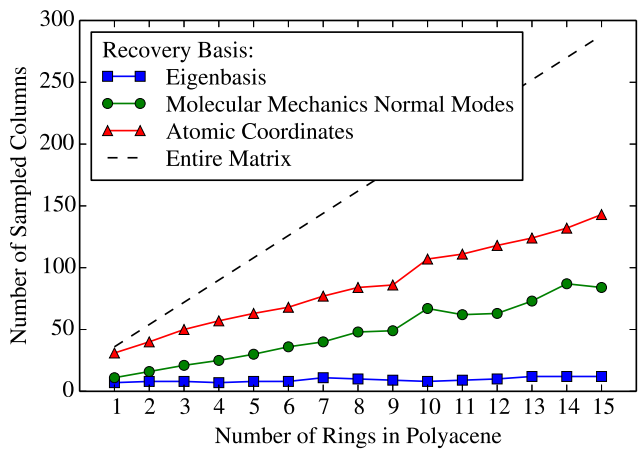}
\caption{\label{fig:NColumnsVsNRings} Number of columns which must be sampled as a function of the number of rings in the polyacene to achieve a relative Frobenius norm error less than $10^{-3}$ in the recovered quantum mechanical Hessian.  Legend entries indicate the (sparse) recovery basis, and columns are always sampled in the DCT basis with respect to the recovery basis.  (Relative error is measured by averaging over ten different trials which sample different sets of random columns.)}
\end{center}
\end{figure}

As Fig.~\ref{fig:HessSparsity} shows the sparsity \(S\) of a quantum mechanical Hessian does not necessarily scale linearly with the number of atoms \(N\) in the molecule.  Fig.~\ref{fig:NColumnsVsNRings} illustrates the cost of recovering the quantum mechanical Hessians of polyacenes using compressed sensing in a variety of sparse bases. Recovering the Hessian in the atomic coordinate basis already provides a considerable computational advantage over directly computing the entire Hessian. In fact, this curve mirrors the sparsity per column curve for quantum mechanical Hessians in Fig.~\ref{fig:HessSparsity}, consistent with our prediction that the number of sampled columns scales as \(O(S/N \log N) \times OE\).  More significantly, recovering the quantum mechanical Hessian in the molecular mechanics basis provides a substantial advantage over recovery in the atomic coordinates basis, reducing the number of columns which must be sampled approximately by a factor of two.  This is consistent with the quantum mechanical Hessian being considerably sparser in the approximate eigenbasis of molecular mechanics normal modes. Of course, nothing beats recovery in the exact eigenbasis, which is as sparse as possible, but which requires knowing the exact Hessian in the first place. 

In short, the take-home message of Fig.~\ref{fig:NColumnsVsNRings} is that using compressed sensing to recover a quantum mechanical Hessian in its basis of molecular mechanics normal modes is a practical procedure which substantially reduces the computational cost of the procedure.

\section{Conclusions}

We have presented a new approach for calculating matrices. This method is suitable for applications where the cost of computing each matrix element is high in comparison to the cost of linear algebra operations.  Our approach leverages the power of compressed sensing to avoid individually computing every matrix element, thereby achieving substantial computational savings. 

When applied to molecular vibrations of organic molecules, our method results in accurate frequencies and normal modes with about 30\% of the expensive quantum mechanical computations usually required, which represents a quite significant 3\(\times\) speed-up. Depending on the sparsity of the Hessian, our method can also improve the overall scaling of the computation. These computational savings could be further improved by using more sophisticated compressed sensing approaches, such as recovery algorithms based on belief propagation\cite{Krzakala_2012a,Krzakala_2012b} which offer a recovery cost directly proportional to the sparsity of the signal, and which could be easily integrated into our approach.

Our method could also be applied to other common calculations in computational chemistry, including the Fock matrix in electronic structure or the Casida matrix in linear-response time-dependent DFT~\cite{TDDFT2012}. Nevertheless, our method is not restricted to quantum chemistry and is applicable to many problems throughout the physical sciences and beyond. The main requirement is an \emph{a priori} guess of a basis in which the matrix to be computed is sparse.  The optimal way to achieve this requirement is problem-dependent, but as research into sparsifying transformations continues to develop, we believe our method will enable considerable computational savings in a wide array of scientific fields.

In fact, a recent area of interest in compressed sensing is the development of dictionary learning methods that do not directly require knowledge of a sparsifying basis, but instead generate it on-the-fly based on the problem~\cite{Aharon_2006,Rubinstein_2010}.  We believe that combining our matrix recovery protocol with state-of-the-art dictionary learning methods may eventually result in further progress towards the calculation of scientific matrices.

Beyond the specific problem of computing matrices, this work demonstrates that compressed sensing can be integrated into the core of computational simulations as a workhorse to reduce numerical costs by optimizing the information obtained from each computation.

Finally, we introduced an effective method of bootstraping low-accuracy calculations to reduce the number of high-accuracy calculations that need to be done, something which is not simple to do in quantum chemical calculations. In this new paradigm, the role of expensive high-accuracy methods is to correct the low-accuracy results, with a cost proportional to the magnitude of the required correction, rather than recalculating the results from scratch.

\section{Computational methods}

The main computational task required to implement our approach is the solution of the \(\ell_1\) optimization problem in eq.~\ref{eq:bpdn}. From the many algorithms available for this purpose, we rely on the spectral projected gradient \(\ell_1\) (SPGL1) algorithm developed by van~den~Berg and Friedlander~\cite{Berg2008} and their freely-available implementation.  

For all compressed sensing calculations in this paper, the change-of-basis matrix between the sparse basis and the measurement basis is given by the DCT matrix whose elements are given by
\begin{align}
\label{eq:dct}
P_{ij} = \sqrt{\dfrac{2}{N}} \cos \left[ \frac{\pi}{N} \left(i-1\right) \left(j - \frac{1}{2}\right) \right] \ ,
\end{align}
with the first row multiplied by an extra factor of \(1/\sqrt{2}\) to guarantee orthogonality.

For the numerical calculations we avoid explicitly constructing the Kronecker product of \(P\) with itself and instead perform all matrix multiplications in the SPGL1 algorithm directly in terms of \(P\).  This latter approach has much smaller memory requirements and numerical costs, ensuring that the compressed sensing process itself is rapid and not a bottleneck in our procedure.  The condition \(PAP^T = B\) is satisfied up to a relative error of \(10^{-7}\) in the Frobenius norm (vectorial 2-norm).

In order to perform the undersampling required for our compressed sensing calculations, first the complete Hessians were calculated, then they were converted to the measurement basis, and finally they were randomly sampled by column.  Quantum mechanical Hessians were obtained with the QChem 4.2~\cite{Shao_2014} software package, using density functional theory with the B3LYP exchange-correlation functional~\cite{Becke_1993} and the 6-31G* basis set.  Molecular mechanics Hessians were calculated using the the MM3 force field~\cite{Allinger_1989} and the open-source package Tinker 6.2.

\begin{acknowledgments}
We acknowledge D. Rappoport and T. Markovich for useful discussions. The computations in this paper were run on the Odyssey cluster supported by the Faculty of Arts and Sciences (FAS) Science Division Research Computing Group at Harvard University. This work was supported by the Defense Threat Reduction Agency under contract no. HDTRA1-10-1-0046, by the Defense Advanced Research Projects Agency under award no. N66001-10-1-4060, and by Nvidia Corporation through the CUDA Center of Excellence program. J.N.S. acknowledges support from the Department of Defense (DoD) through the National Defense Science and Engineering Graduate Fellowship (NDSEG). 
\end{acknowledgments}

\bibliography{biblio}

\end{document}